%% file: TOP-16-001_temp.tex
\pdfoutput=1

\documentclass[11pt,twoside,a4paper,cmspaper,final,collab]{cms-tdr}
\def\svnVersion{396011}\def\cmsCernNoTag{CERN-EP/2016-266}\def\cmsCernDate{\today}\def\cmsMessage{Published in the Journal of High Energy Physics as \href{http://dx.doi.org/10.1007/JHEP03(2017)101}{\doi{10.1007/JHEP03(2017)101.}}}

\begin{document}\cmsNoteHeader{TOP-16-001}

\hyphenation{had-ron-i-za-tion}
\hyphenation{cal-or-i-me-ter}
\hyphenation{de-vices}
\RCS$Revision: 386692 $
\RCS$HeadURL: svn+ssh://svn.cern.ch/reps/tdr2/papers/TOP-16-001/trunk/TOP-16-001.tex $
\RCS$Id: TOP-16-001.tex 386692 2017-02-12 18:50:20Z alverson $

\newcommand{\ACP}{\ensuremath{A_\mathrm{CP}}\xspace}
\newcommand{\ACPp}{\ensuremath{A_\mathrm{CP}^\prime}\xspace}
\newcommand{\Otwo}{\ensuremath{O_{2}}\xspace}
\newcommand{\Othree}{\ensuremath{O_{3}}\xspace}
\newcommand{\Ofour}{\ensuremath{O_{4}}\xspace}
\newcommand{\Oseven}{\ensuremath{O_{7}}\xspace}

\cmsNoteHeader{TOP-16-001}
\title{Search for CP violation in \ttbar production and decay in proton-proton collisions at $\sqrt{s}=8\TeV$}

\date{\today}

\abstract{
The results of a first search for CP violation in the production and decay of top quark-antiquark (\ttbar) pairs are presented. The search is based on asymmetries in T-odd, triple-product correlation observables, where T is the time-reversal operator. The analysis uses a sample of proton-proton collisions at $\sqrt{s}=8\TeV$ collected by the CMS experiment, corresponding to an integrated luminosity of 19.7\fbinv. Events are selected having one electron or muon and at least four jets. The T-odd observables are measured using four-momentum vectors associated with \ttbar production and decay. The measured asymmetries exhibit no evidence for CP-violating effects, consistent with the expectation from the standard model.}

\hypersetup{%
pdfauthor={CMS Collaboration},%
pdftitle={Search for CP violation in top quark-antiquark production and decay in proton-proton collisions at sqrt(s) = 8 TeV},%
pdfsubject={CMS},%
pdfkeywords={CMS, physics, top, CP violation}}

\maketitle

\section{Introduction}\label{sec:introduction}

Violation of the combined operation of charge conjugation and parity (CP) is introduced in the standard model (SM) via an irreducible phase in the Cabibbo--Kobayashi--Maskawa quark-mixing matrix~\cite{Hocker:2006xb}. Detailed experimental investigation of CP violation (CPV) in the strange and bottom quark sectors has been conducted over the past few decades~\cite{Bevan:2014iga}. The measured asymmetries are well described by the SM, but are too small to explain the observed matter-antimatter asymmetry of the universe~\cite{Agashe:2014kda}. In contrast to the strange and bottom quark sectors, CPV in the top quark sector is relatively unexplored.
In the SM, CPV in the production and decay of top quark-antiquark (\ttbar) pairs is predicted to be very small~\cite{CPVtop:General}.
However, in many theories of physics beyond the SM (see, for example, Refs.~\cite{CPVtop:Tevatron,CPVtop:LHCDilepton} and references therein) sizable CP-violating effects could be observed, which have the potential to shed light on the matter-antimatter asymmetry of the universe.

In this paper, the first measurements of CP-violating asymmetries in \ttbar production and decay are presented. One of the top quarks is presumed to decay to a bottom (b) quark and a hadronically decaying {\PW} boson. The other top quark is required to decay to a b quark and a {\PW} boson that decays leptonically to an electron or muon and its associated neutrino. The analysis exploits T-odd, triple-product correlations, where T is the time-reversal operator. Several observables are measured, as proposed in Refs.~\cite{CPVtop:Tevatron,CPVtop:LHCDilepton,CPVtop:LHCLeptonJets}, that take the form $\vec{v}_1\cdot (\vec{v}_2\times \vec{v}_3)$, where $\vec{v}_i$ $(i=1,2,3)$ are spin or momentum vectors. These triple-product observables are odd under the T transformation, and are thus also odd under the CP transformation if CPT conservation is valid, \ie $\mathrm{CP}(O_{i}) = -O_{i}$, where $O_i$ are the proposed observables. The presence of CPV would be manifested by a nonzero value of the asymmetry

\begin{equation}\label{eq:acp}
    A_\mathrm{CP}(O_{i})=\frac{N_\text{events}(O_{i}>0)-N_\text{events}(O_{i}<0)}{N_\text{events}(O_{i}>0)+N_\text{events}(O_{i}<0)}\ .
\end{equation}

The measurements of the asymmetry corrected for the effects of the detector (\ACP) and also without these corrections (\ACPp) are presented. The reason to present both \ACP and \ACPp values is that the corrections, called dilution factors (Section~\ref{ss:dilutionfactor_results}), could themselves be affected by physics beyond the SM~\cite{CPVtop:LHCLeptonJets}; no particular such new-physics process is considered in this paper.

Four observables that can be measured in the single-lepton + jets final state of \ttbar production and decay in proton-proton (\Pp\Pp) collisions are defined as:

\begin{equation}\label{eq:defO}
    \begin{split}
        &O_{2}=\epsilon(P,p_{\PQb}+p_{\PAQb},p_{\ell},p_{j_{1}})\xrightarrow{\text{lab}}\propto (\vec{p}_{\PQb}+\vec{p}_{\PAQb})\cdot(\vec{p}_{\ell}\times\vec{p}_{j_{1}}),\\
        &O_{3}=Q_{\ell}\,\epsilon(p_{\PQb},p_{\PAQb},p_{\ell},p_{j_{1}})\xrightarrow{\bbbar\, \mathrm{CM}} \propto Q_{\ell}\,\vec{p}_{\PQb}\cdot(\vec{p}_{\ell}\times\vec{p}_{j_{1}}),\\
        &O_{4}=Q_{\ell}\,\epsilon(P,p_{\PQb}-p_{\PAQb},p_{\ell},p_{j_{1}})\xrightarrow{\text{lab}} \propto Q_{\ell}\,(\vec{p}_{\PQb}-\vec{p}_{\PAQb})\cdot(\vec{p}_{\ell}\times\vec{p}_{j_{1}}),\\
        &O_{7}=q\cdot(p_{\PQb}-p_{\PAQb})\,\epsilon(P,q,p_{\PQb},p_{\PAQb})\xrightarrow{\text{lab}} \propto(\vec{p}_{\PQb}-\vec{p}_{\PAQb})_{z}(\vec{p}_{\PQb}\times\vec{p}_{\PAQb})_{z}.
    \end{split}
\end{equation}

The symbol $\rightarrow$ indicates the spatial frame chosen to simplify the triple product. The observables \Otwo, \Ofour, and \Oseven are calculated in the laboratory (lab) frame, and \Othree in the \bbbar centre-of-mass frame (\bbbar CM), where \PQb and \PAQb indicate the bottom quark and antiquark jets from the \PQt and \PAQt decays, respectively.
 The symbol $\propto$ indicates proportionality.
 The symbol $\epsilon$ denotes the Levi--Civita symbol with $\epsilon_{0123}=1$, which is contracted with four-vectors $a$, $b$, $c$, and $d$, \ie $\epsilon(a,b,c,d)\equiv\epsilon_{\mu\nu\alpha\beta}a^{\mu}b^{\nu}c^{\alpha}d^{\beta}$. In these expressions, $P$ is the sum of, and $q$ the difference between, the four-momenta of the two initial-state protons; $p$ and $\vec{p}$ are the four- and three-momenta, respectively, of the final-state particles; the subscript $z$ indicates a projection along the direction of the counterclockwise rotating proton beam, defined to be the $+z$ direction in the CMS coordinate system; $\ell$ refers to the electron or muon from the leptonically decaying {\PW} boson; $j_1$ refers to the non-b quark jet originating from the hadronically decaying {\PW} boson with the highest transverse momentum (\pt); and $Q_{\ell}$ is the electric charge of $\ell$.
 Note that the sign of the observable is the only information needed to measure \ACP.

The asymmetries \ACP computed from the above observables are predicted to be zero in the SM~\cite{CPVtop:Tevatron,CPVtop:LHCDilepton}. However, in some new-physics scenarios~\cite{CPVtop:LHCLeptonJets}, the effects of CPV can be sizable: $\ACP(\Othree)$ and $\ACP(\Ofour)$ could be as large as 8\%, while $\ACP(\Otwo)$ and $\ACP(\Oseven)$ are less sensitive to new physics and can reach 0.4\%~\cite{CPVtop:LHCLeptonJets}. The sensitivity of the observables to CPV depends on whether distinguishable final-state objects are involved in their definition. For instance, the \PQb~quark jet charges need to be distinguished for \Othree and \Ofour, but not for \Otwo and \Oseven.

\section{The CMS detector}\label{sec:detector}

The central feature of the CMS apparatus is a superconducting solenoid of 6\unit{m} internal diameter, providing a magnetic field of 3.8\unit{T}. Within the solenoid volume are a silicon pixel and strip tracker, a lead tungstate crystal electromagnetic calorimeter (ECAL), and a brass and scintillator hadron calorimeter, each composed of a barrel and two endcap sections. Forward calorimeters extend the pseudorapidity ($\eta$) coverage provided by the barrel and endcap detectors. Muons are measured in gas-ionization detectors embedded in the steel flux-return yoke outside the solenoid.
A more detailed description of the CMS detector, together with a definition of the coordinate system and relevant kinematic variables, is given in Ref.~\cite{Chatrchyan:2008aa}.

\section{Data and simulated samples}\label{sec:samples}

This analysis uses data from $\sqrt{s}=8\TeV$ \Pp\Pp~collisions collected with the CMS detector in 2012, corresponding to an integrated luminosity of 19.7\fbinv.

Monte Carlo (MC) simulations are used to model the SM processes of relevance for this analysis.
 Top quark-antiquark events are generated at leading order using the \MADGRAPH (v5.1.3.30) program~\cite{MadGraph} with the CT10~\cite{PDF:CT10NNLO} parton distribution functions (PDFs). The \MADGRAPH generator accounts for the spin correlations between the top quark and antiquark. The mass of the top quark in the simulation is set to $m_{\PQt}=172.5\GeV$. The \MADGRAPH generator is interfaced with the \PYTHIA (v6.426) generator~\cite{Sjostrand:2006za} with Tune Z2*~\cite{Chatrchyan:2013gfi} to simulate parton showering and hadronization.
 The \ttbar production cross section is calculated with the \textsc{Top++} 2.0 package~\cite{Cacciari:2011hy} and estimated up to next-to-next-to-leading order (NNLO).
 It has been observed in CMS measurements~\cite{Khachatryan:2015oqa,Khachatryan:2015fwh} that \MADGRAPH exhibits a mismodelling of the top quark \pt in \ttbar events.
 To rectify this deficiency, an additional top quark \pt reweighting is applied at generator level to obtain agreement in the \pt spectra between data and simulation.
 The weighting factors are derived from the results of Ref.~\cite{Khachatryan:2015oqa}.

Several background processes are considered in the analysis.
Single top quark production is the main background, and is simulated with the \POWHEG (v1.0) program~\cite{POWHEG,Nason:2004rx,Alioli:2010xd,Frixione:2007vw,Alioli:2009je,Re:2010bp}. The cross section is calculated with the \textsc{Hathor} (v2.1) program~\cite{HATHOR:SingelTop,HATHOR:HArdronTop} at next-to-leading order (NLO).
Drell--Yan (DY) and W + jets processes are generated with \MADGRAPH, and diboson events (WW, WZ, and ZZ) with \PYTHIA.
The cross sections of W + jet and DY production are calculated with the \FEWZ (v3.1)~\cite{Gavin:2012sy,Gavin:2010az} program at NNLO, while the \MCFM (v6.6)~\cite{Campbell:2010ff} program is used for the cross sections of diboson productions at NLO. The quantum chromodynamic (QCD) background is suppressed by the selection requirements (Section~\ref{sec:evtSelection}) and is negligible in the signal region.

All generated events are subjected to a \GEANTfour-based~\cite{geant} simulation of the detector response.
 Additional \Pp\Pp~interactions occurring in the same or nearby bunch crossing (pileup) are included in the simulation. The number of pileup events in simulation is corrected to agree with data. The average number of interactions per event in data is 21.

\section{Object definition and event selection}\label{sec:evtSelection}

The event selection is based on the signature of the single-lepton + jets decay channel of the \ttbar process. Events containing one isolated electron or muon and at least four jets, including exactly two jets identified as originating from the hadronization of b quarks (b quark jets), are selected. Events with electrons and muons coming from the decay of $\tau$ leptons are included in the sample. Electrons, muons, photons, and neutral and charged hadrons are reconstructed and identified with the CMS particle-flow (PF) algorithm \cite{CMS-PAS-PFT-09-001,CMS-PAS-PFT-10-001}.
 The energy of electrons is determined from a combination of the track momentum at the primary collision vertex, the corresponding cluster of energy deposits in the ECAL, and the energy sum of all bremsstrahlung photons associated with the track \cite{Khachatryan:2015hwa}.
 The momentum of muons is obtained from a global fit to signals registered by the silicon tracker and muon detectors~\cite{MuonPerfomance}. The trigger requires at least one electron candidate with $\pt > 27\GeV$ and $\abs{\eta} < 2.5$, or at least one muon candidate with $\pt > 24\GeV$ and $\abs{\eta} < 2.1$.
 The primary event vertex is the reconstructed vertex with the largest $\pt^2$ sum of associated tracks.

In the subsequent offline selection, electrons are identified using a combination of the ECAL shower shape information, and the geometric matching between the track and the ECAL energy cluster~\cite{Khachatryan:2015hwa}. The electron candidates are required to have $\pt>30\GeV$ and $\abs{\eta} < 2.1$, excluding the transition region ($1.44 < |\eta| < 1.56$) between the barrel and endcap sections of the ECAL, where the acceptance is difficult to accurately model. Electrons from identified photon conversions are rejected. Muon candidates must be associated with a good-quality track~\cite{MuonPerfomance} with $\pt > 26\GeV$ and $\abs{\eta} < 2.1$. The trigger and lepton identification efficiencies are determined from data and simulation~\cite{Khachatryan:2010xn} as functions of the lepton \pt and $\eta$. The electron (muon) identification efficiency ranges from 55 to 85\% (93 to 97\%).

The lepton from the W boson decay is expected to be isolated from other activity in the event. A relative isolation parameter~\cite{MuonPerfomance,Khachatryan:2015hwa} is defined as the scalar \pt sum of the particles reconstructed by the PF algorithm within a cone of angular radius $\Delta R = \sqrt{\smash[b]{(\Delta\eta)^{2} + (\Delta\phi)^{2} }} = 0.3$\,(0.4) for electrons (muons) around the lepton candidate direction, divided by the \pt of the lepton candidate,
where $\Delta\eta$ and $\Delta\phi$ are the relative differences in pseudorapidity and azimuthal angle (in radians), respectively, between the directions of the lepton and other particle.
The sum includes pileup corrections and excludes the lepton candidate under consideration. The relative isolation parameter is required to be less than 0.10 for electrons and 0.12 for muons. Events with additional loosely defined leptons (satisfying a lower-\pt threshold and a less stringent isolation requirement) are rejected in order to reduce the contributions from Z boson or \ttbar decays into dileptons.

Jets are reconstructed by clustering charged and neutral PF particles, using the anti-\kt jet algorithm~\cite{Cacciari:2008gp} with a distance parameter of 0.5, implemented in the $\textsc{FastJet}$ package~\cite{Cacciari:2011ma}.
Jet energies are corrected for the nonlinear response of the calorimeters and for the differences between the measured and simulated responses~\cite{Khachatryan:2016kdb}. Charged hadrons that are not associated with the primary vertex are removed, and the jet energy is corrected to account for the expected contributions of neutral particles from pileup interactions~\cite{CMS-PAS-PFT-09-001,CMS-PAS-PFT-10-002}. An event is discarded if the lepton candidate lies within $\Delta R = 0.5$ of any selected jet.

At least four jets with $\pt > 30\GeV$ and $\abs{\eta} < 2.4$ are required, two of which must be identified as b quark jets. The b tagging is performed with the combined secondary vertex (CSV) algorithm at the medium working point~\cite{Chatrchyan:2012jua}, corresponding to an efficiency of about $1\%$ for light-quark and gluon jets (mistag rate) and 60--70\% for b quark jets, depending on the jet \pt and $\eta$. The MC simulation is corrected with scale factors to account for differences with respect to the data for the b tagging efficiency~\cite{CMS:2013vea}.

\begin{figure}[htb]
  \centering
    \includegraphics[width=0.45\textwidth]{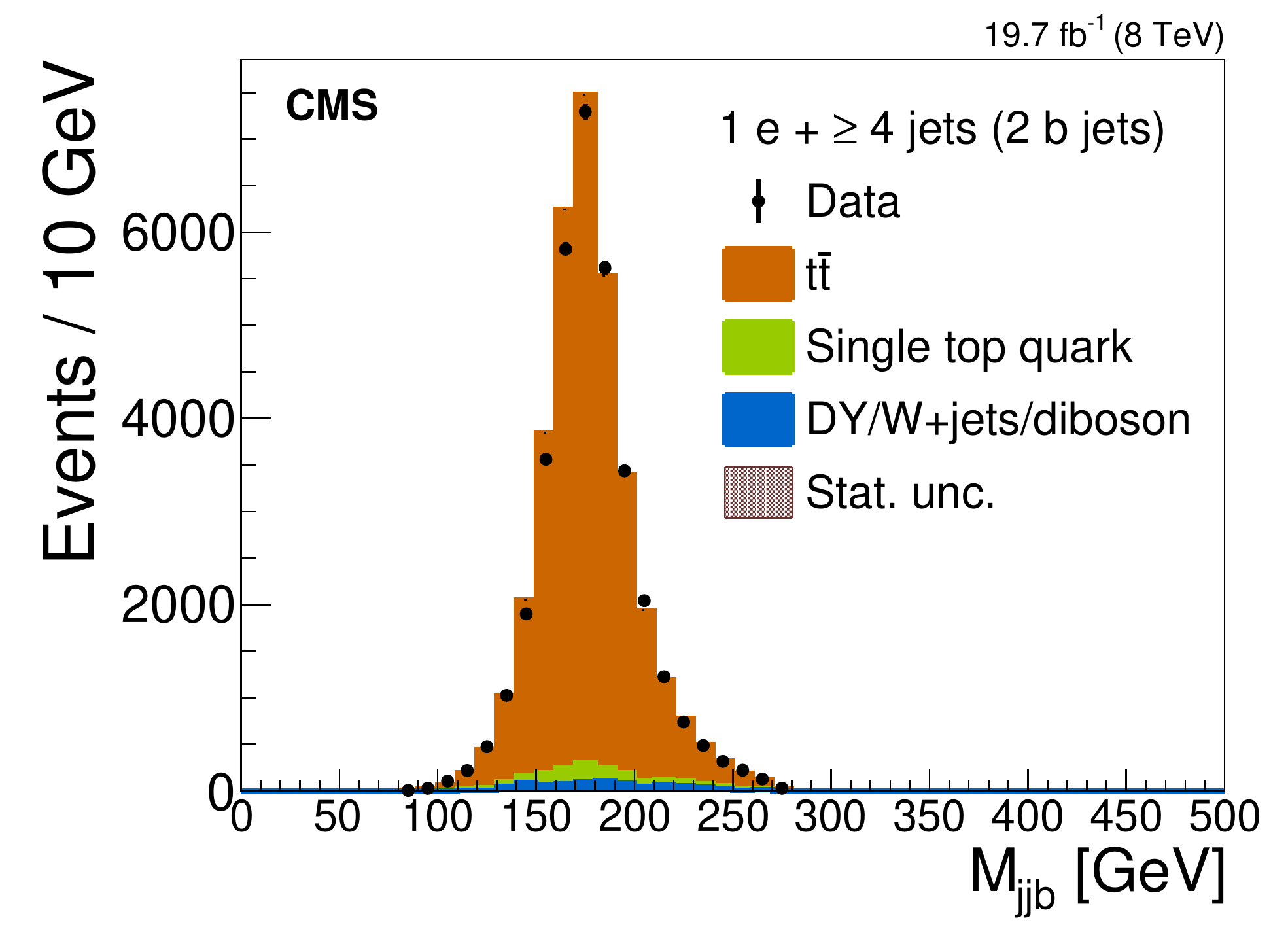}\label{fig:evtSelction_SR:El_hmass}
    \includegraphics[width=0.45\textwidth]{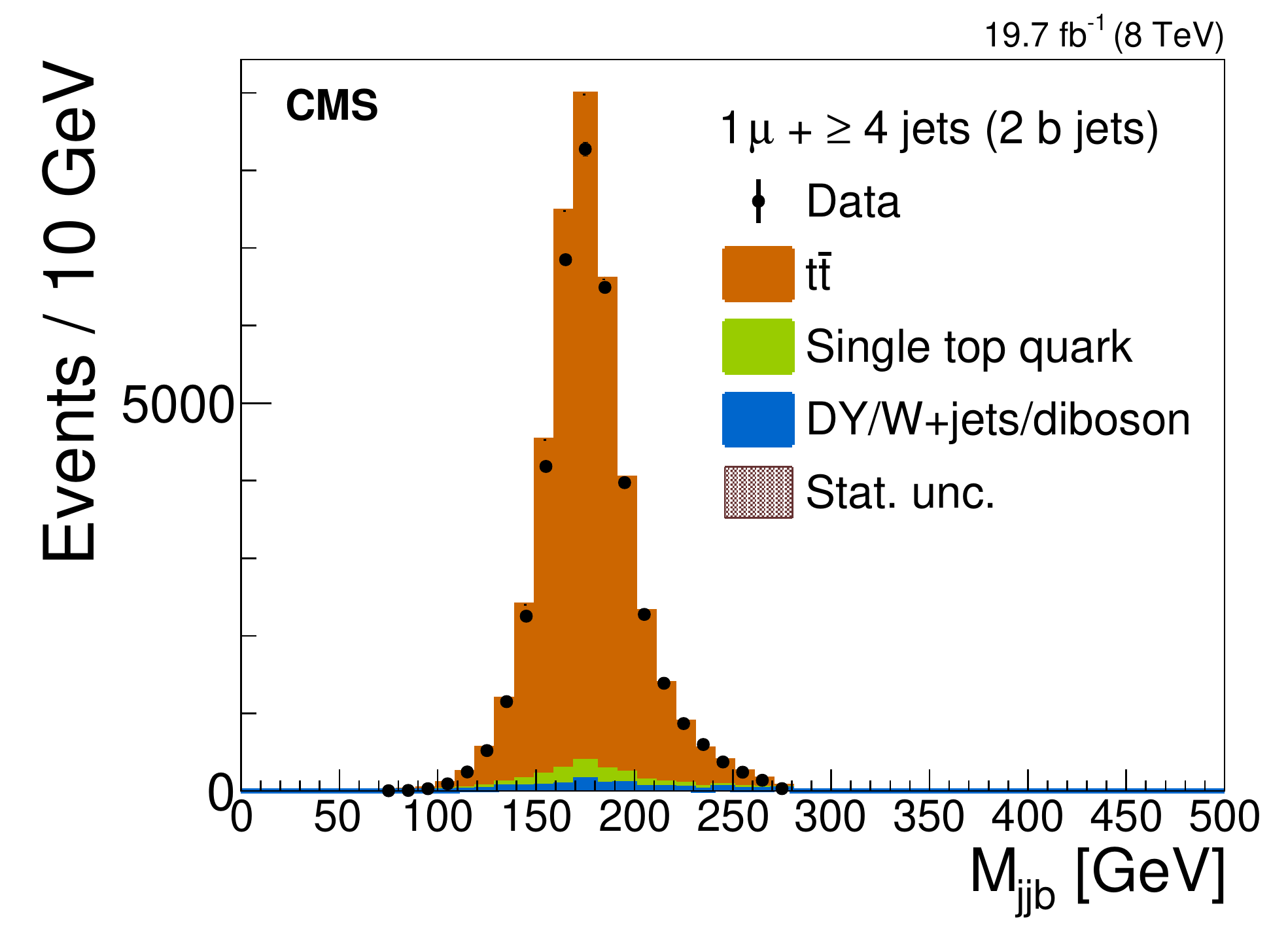}\label{fig:evtSelction_SR:Mu_hmass} \\
    \includegraphics[width=0.45\textwidth]{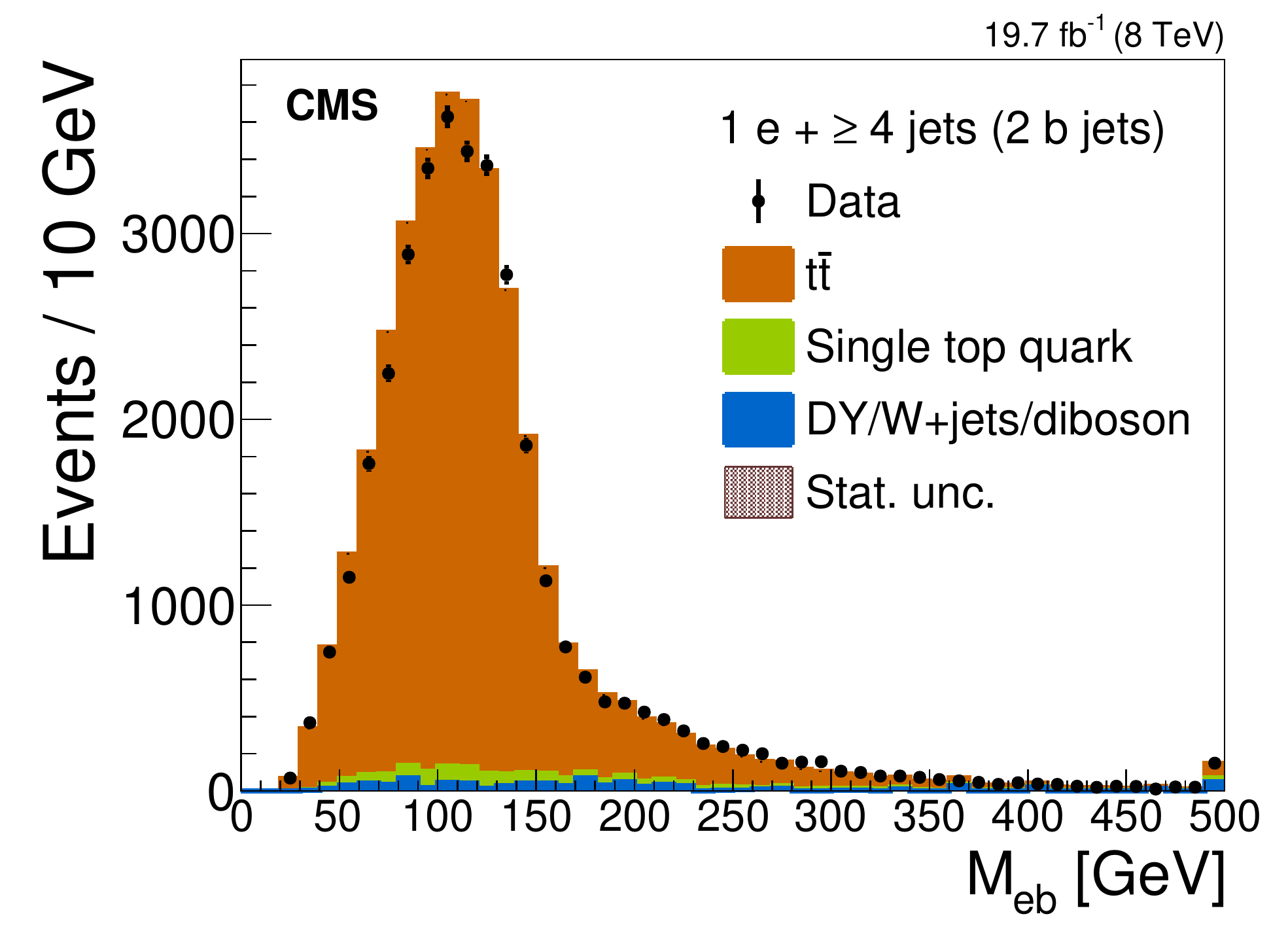}\label{fig:evtSelction_SR:El_lmass}
    \includegraphics[width=0.45\textwidth]{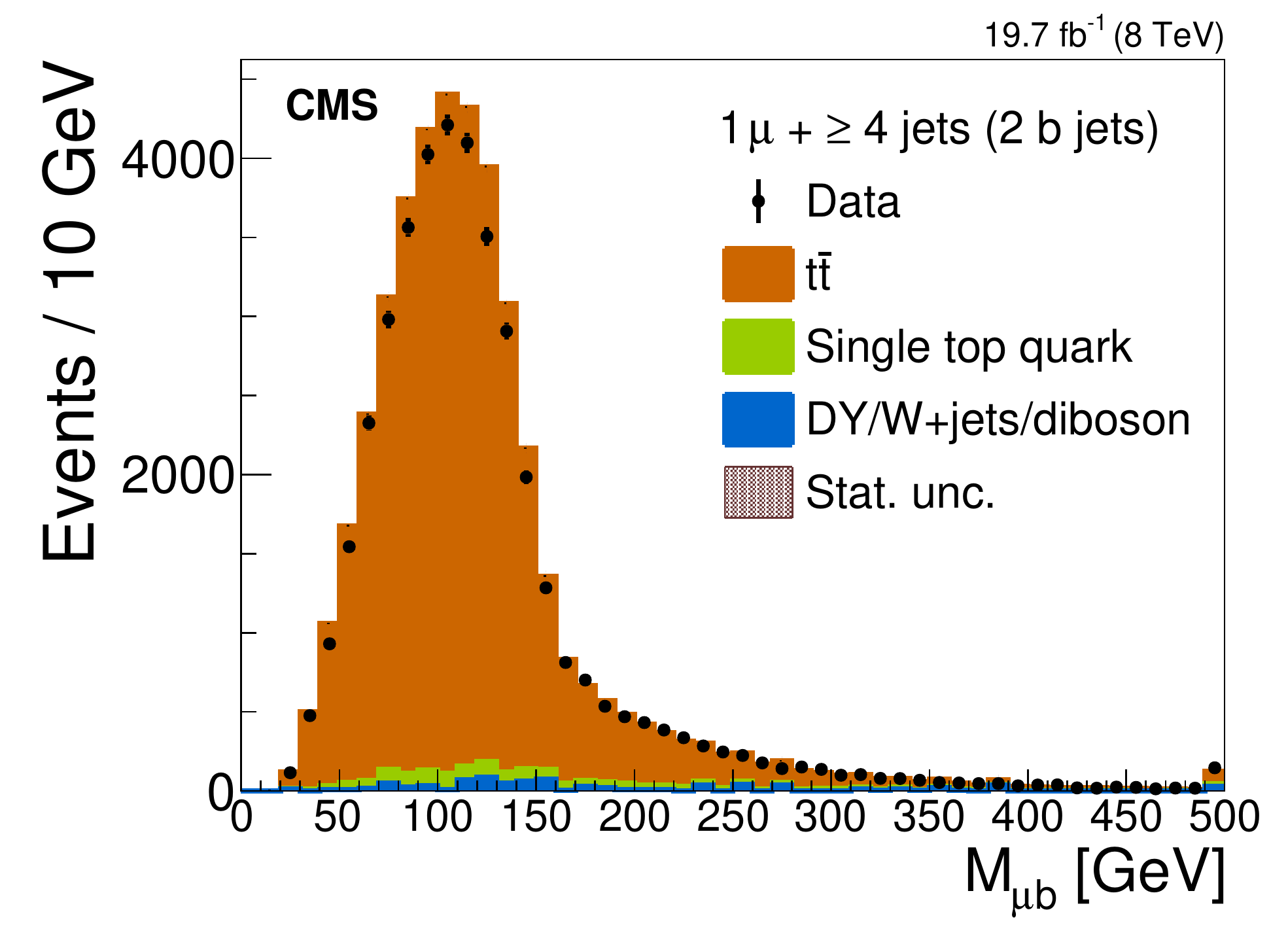}\label{fig:evtSelction_SR:Mu_lmass}
  \caption{The measured invariant mass distributions from data (points) of (upper) hadronically and (lower) semileptonically decaying top quark candidates in the (left) electron and (right) muon channels, compared to the predictions for the signal and various backgrounds from simulation (filled histograms). The QCD background is found to be negligible. The overflow events are collected in the last bins. The vertical bars on the data points and the hatched bands indicate the statistical uncertainties in the data and simulation, respectively.}
    \label{fig:evtSelction_SR}
\end{figure}

One of the b quark jets is combined with two non-b quark jets in the event through a $\chi^2$-sorting algorithm~\cite{Chatrchyan:2012jua,Khachatryan:2016yzq} that makes use of top quark and {\PW} boson mass constraints to define the hadronically decaying top quark candidate.
 The other \PQb~quark jet is then associated with the semileptonically decaying top quark candidate.
 The purity of the selected \ttbar candidates is $\approx$92\% after the $\chi^2$ selection, while single top quark production contributes with only $\approx$3\% of the total number of events.
 The charge of the isolated lepton is used to distinguish between the b and \PAQb quark jets.
 The b quark jet charges are correctly assigned in $\approx$60\% of the \ttbar events.
 For the semileptonically decaying top quark, the analysis uses the mass variable $M_{\ell\PQb}$, which is the invariant mass of the isolated lepton and the associated b quark jet. For the hadronically decaying top quark, the mass variable $M_{jj\PQb}$ is used, which is the invariant mass of the two non-b quark jets and the associated b quark jet.
 As shown in Fig.~\ref{fig:evtSelction_SR}, although the analysis does not depend on the simulated background events, there is reasonable agreement in the two mass-variable distributions between the data and the simulation.
 The distributions of the two mass variables for data events with positive and negative $O_{i}$ values are consistent with each other in both the electron and muon channels.

\section{Background control sample and check for asymmetry bias}\label{sec:control_region}

Detector and reconstruction effects may induce spurious results for the asymmetries. A data-control sample (CS) enriched in random combinations of leptons and jets (combinatorial background) from non-\ttbar events is used to check for these spurious effects and to evaluate possible bias, \ie a nonzero measured value for \ACPp. The combinatorial background is assumed to possess no intrinsic CPV because of its random nature.

To enhance the fraction of background events and minimize the contribution from \ttbar signal events, the CS is selected by requiring no b-tagged jets defined according to the loose working point of CSV~\cite{Chatrchyan:2012jua}, which is 80\% efficient in identifying b jets with a mistag rate for lighter jets of 10\%.
 The event is rejected if there are additional nonisolated electron or muon candidates. All other event selection requirements are equivalent to the signal region. The CS is expected to be dominated by non-\ttbar processes $\approx$90\%), with the major contribution from the W + jets process. The CS contains a sufficiently large number of events to perform statistically significant cross-checks. The kinematic distributions from data in the CS are in agreement with those of the background simulation in the signal region, and thus the CS can be used to represent the signal-region background.

Since the CS does not contain any events with a tagged b quark jet, the two jets with the highest CSV discriminator values are used to play the role of the b quark jets. Using the same procedure as for the signal region, \ie the $\chi^2$--sorting algorithm, the required objects are found and the \ACPp values are determined. The CS asymmetry measurements are consistent with zero within about two standard deviations of the statistical uncertainty in both the electron and muon channels, as shown in Table~\ref{tab:ACP_0bCR}. The combined electron and muon \ACPp results are determined by summing the distributions in the electron and muon channels after applying the background normalization from the fit described in Section~\ref{sec:bkgEstimation}. The systematic uncertainties given in Table~\ref{tab:ACP_0bCR} are derived from the fit results described in Section~\ref{sec:systematics}. The distributions of the observables of interest measured in the CS are used for background subtraction in the signal region.

\begin{table}[tb]
 \centering
  \topcaption{
 The uncorrected CP asymmetry \ACPp, measured in percent, obtained from the control sample as described in the text for each of the four observables. Results are given for the electron and muon channels separately and for their combination. For the separate electron and muon channels, the uncertainties are statistical. For the combined results, the first uncertainty is statistical and the second systematic.}
  \label{tab:ACP_0bCR}
  \begin{tabular}{cccc}
   \multicolumn{4}{c}{$\ACPp\mathrm{(CS)}$ (\%)} \\
    &  e + jets & $\mu$ + jets & $\ell$ + jets \\
   \hline
   $\Otwo$    & $+0.30 \pm 0.59$ & $-0.67 \pm 0.65$ & $-0.23 \pm 0.44 \pm 0.02 $\\
   $\Othree$  & $+0.85 \pm 0.59$ & $-0.57 \pm 0.65$ & $+0.08 \pm 0.44 \pm 0.03 $\\
   $\Ofour$   & $+1.19 \pm 0.59$ & $-0.42 \pm 0.65$ & $+0.32 \pm 0.44 \pm 0.04 $\\
   $\Oseven$  & $-0.10 \pm 0.59$ & $+0.60 \pm 0.65$ & $+0.28 \pm 0.44 \pm 0.02 $\\
  \end{tabular}
\end{table}
\section{Fit procedure and \texorpdfstring{$A'_\mathrm{CP}$}{A'[CP]} determination}\label{sec:bkgEstimation}

The yields of signal and background events are extracted by maximizing an extended likelihood function from the data, using the $M_{\ell\PQb}$ distribution. The expected signal distribution is obtained from simulation, while the background distribution is obtained from the CS.

\begin{figure}[htb]
  \centering
    \includegraphics[width=0.49\textwidth]{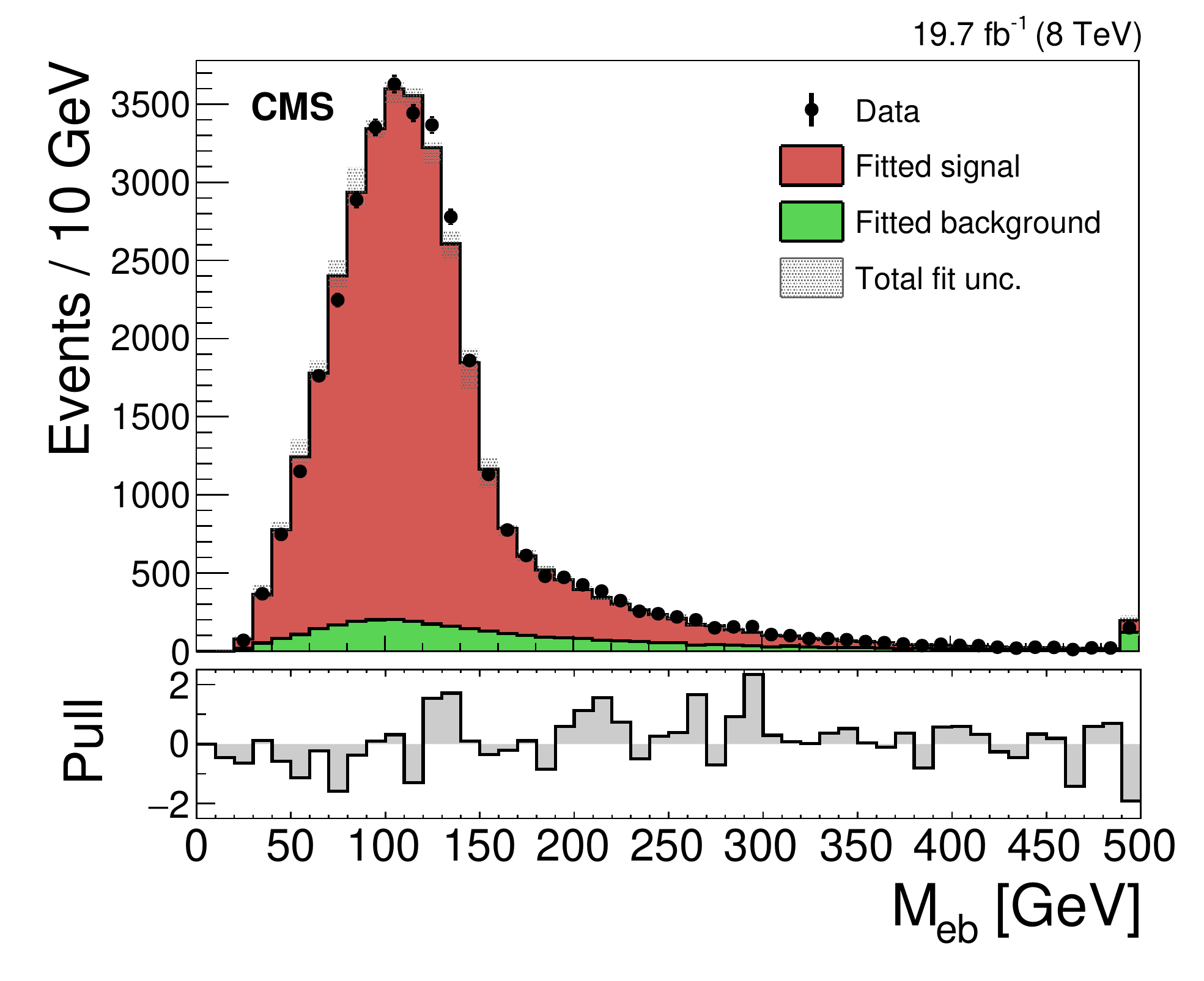}
    \includegraphics[width=0.49\textwidth]{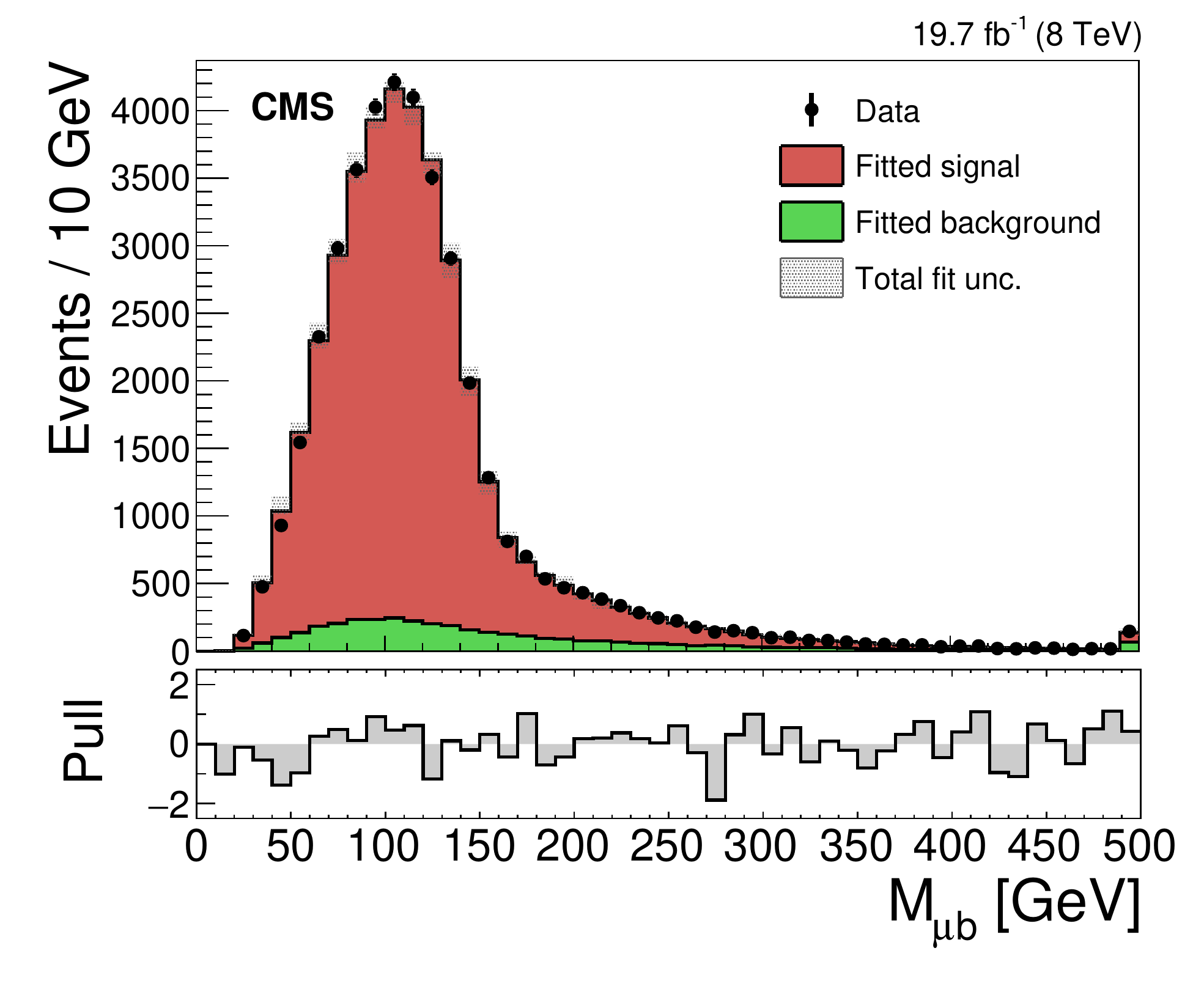}
    \caption{Distribution of the invariant mass $M_{\ell\PQb}$ of the semileptonically decaying top quark candidates for the (left) electron and (right) muon channels, in comparison to the results of the fit described in the text. Overflow events are collected in the last bins. The vertical bars on the data points indicate the statistical uncertainties. The hatched bands shows the combined statistical and systematic uncertainties in the fit results added in quadrature. The difference between the observed and fitted events, divided by the total statistical and systematic uncertainty (pull), is shown for each bin in the lower panels.}
    \label{fig:Fitted}
\end{figure}

Figure~\ref{fig:Fitted} shows the measured $M_{\ell\PQb}$ distributions compared to the results of the fit.
The fit is seen to provide a good representation of the data for both the electron and muon channels.
For the final measurement, only events with $M_{\ell\PQb} < 200\GeV$ are considered because events with large $M_{\ell\PQb}$ suffer from a high rate of incorrect lepton and b quark jet assignments, as determined from simulation. Imposing this requirement increases the sensitivity of final measurements to CPV. The final results for the number of events in the electron and muon channels, and the corresponding \ttbar event purities, are presented in Table~\ref{tab:estimatedNum}. The systematic uncertainties are obtained from the differences between the yields under the variations described in Section~\ref{sec:systematics}.

\begin{table}[tb]
 \centering
  \topcaption{
  The observed and fitted number of events in the electron and muon channels as well as the fitted \ttbar fraction (purity) in percent. While the fit is performed over the full mass range, the fitted and observed results are for $M_{\ell\PQb} < 200\GeV$.  The first uncertainty is statistical and the second systematic.
  }
  \label{tab:estimatedNum}
  \begin{tabular}{ccc}
          & e + jets  & $\mu$ + jets \\
   \hline
   Data   & $31\,129$   & $36\,467$ \\
   \hline
   Fitted events    & $31\,280\pm170\pm40$ & $36\,510\pm190\pm50$ \\
   Fitted \ttbar fraction (\%) & $92.5\pm0.5\pm2.3$   & $92.4\pm0.6\pm2.8$ \\
  \end{tabular}
\end{table}

The shape of the background distribution is obtained from the CS, while its yield is estimated from the fit to the $M_{\ell\PQb}$ distribution. The \ACPp results are then computed using an analogous equation to Eq.~(\ref{eq:acp}), after subtracting the estimated background contribution from the measured observables.
\section{Systematic uncertainties}\label{sec:systematics}

Several sources of systematic uncertainty can affect the measurement of \ACPp. The largest uncertainty comes from possible intrinsic detector bias. As shown in Section~\ref{sec:control_region}, the CS results are compatible with no asymmetry. The statistical uncertainties in the asymmetries measured from the CS are 0.59\%, 0.65\%, and 0.44\% for the electron, muon, and combined channels, respectively, as indicated in Table~\ref{tab:ACP_0bCR}. These uncertainties are taken as the systematic uncertainty from possible detector bias.

Other systematic uncertainties, described below, associated with the signal and background yields are evaluated by repeating the $M_{\ell\PQb}$ fit described in Section~6 under different conditions in correction and modelling.
The uncertainty related to pileup modelling is estimated by varying the inelastic \Pp\Pp~cross section in the simulation by $\pm\,5\%$~\cite{Antchev:2011vs}.
The uncertainties from the lepton identification and isolation efficiencies are determined by varying the data-to-simulation scale factors according to their uncertainties. The jet energy scale and resolution are varied according to their $\eta$- and  $\pt$-dependent uncertainties~\cite{CMS-PAS-PFT-09-001,CMS-PAS-PFT-10-002}.
The scale factors used to correct the b tagging identification probabilities are varied according to their uncertainties~\cite{Chatrchyan:2012jua}.
Effects related to the modelling of the PDFs of the initial-state protons are estimated by varying the CT10 nominal prediction by its eigen-uncertainty sources. Each source is used to derive event-by-event weights, which are then applied to obtain a variation of the signal shape. The envelope of the variations is normalized to reflect a 68\% confidence level~\cite{Lai:2010vv}.
 The uncertainties related to the \ttbar simulation are evaluated by varying the matrix-element-to-parton-shower matching thresholds and the factorization and renormalization scales by factors of 2.0 and 0.5 with respect to their nominal values in the simulation. A modelling uncertainty is determined through comparison to a \ttbar sample generated with \POWHEG. The uncertainty related to the modelling of the top quark \pt spectrum in \MADGRAPH is assessed by varying the weight applied by a factor of two. The uncertainty associated with the top quark mass is estimated by repeating the $M_{\ell\PQb}$ fit using different top quark masses. The observed difference is scaled to reflect an uncertainty of $\pm\,1\GeV$ in the mass~\cite{Khachatryan:2015hba}. These systematic uncertainties contribute to the intermediate stages of the analysis but are observed to largely cancel in the measurements of the CP asymmetry. Thus, the dominant systematic uncertainty arises from the statistical uncertainty in the measurements of possible detector bias (Table~\ref{tab:ACP_0bCR}).

\section{Results}\label{sec:results}

\subsection{Experimental sensitivity study}\label{ss:dilutionfactor_results}

\begin{figure}[htb]
  \centering
    \includegraphics[width=0.42\textwidth]{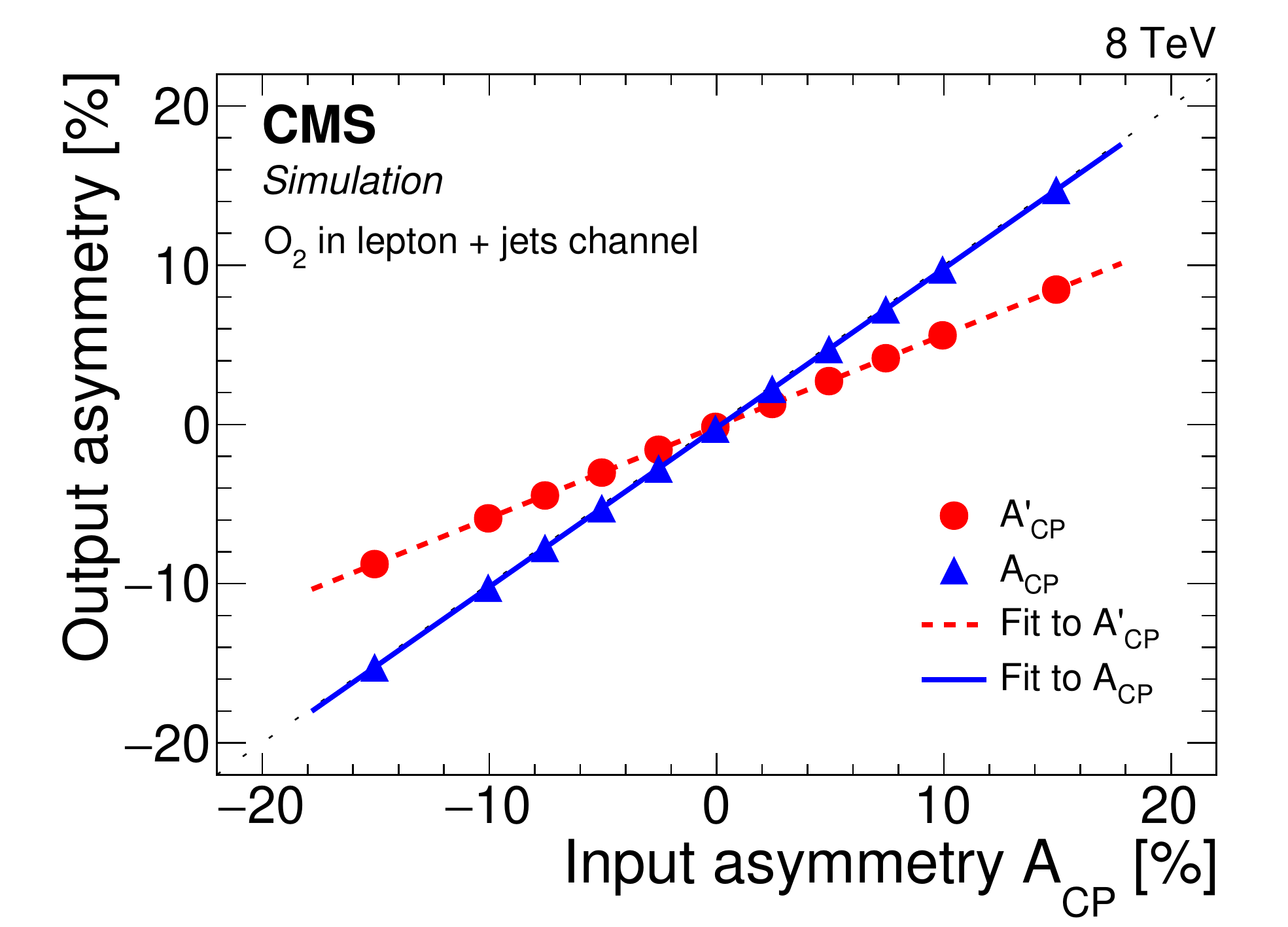}
    \includegraphics[width=0.42\textwidth]{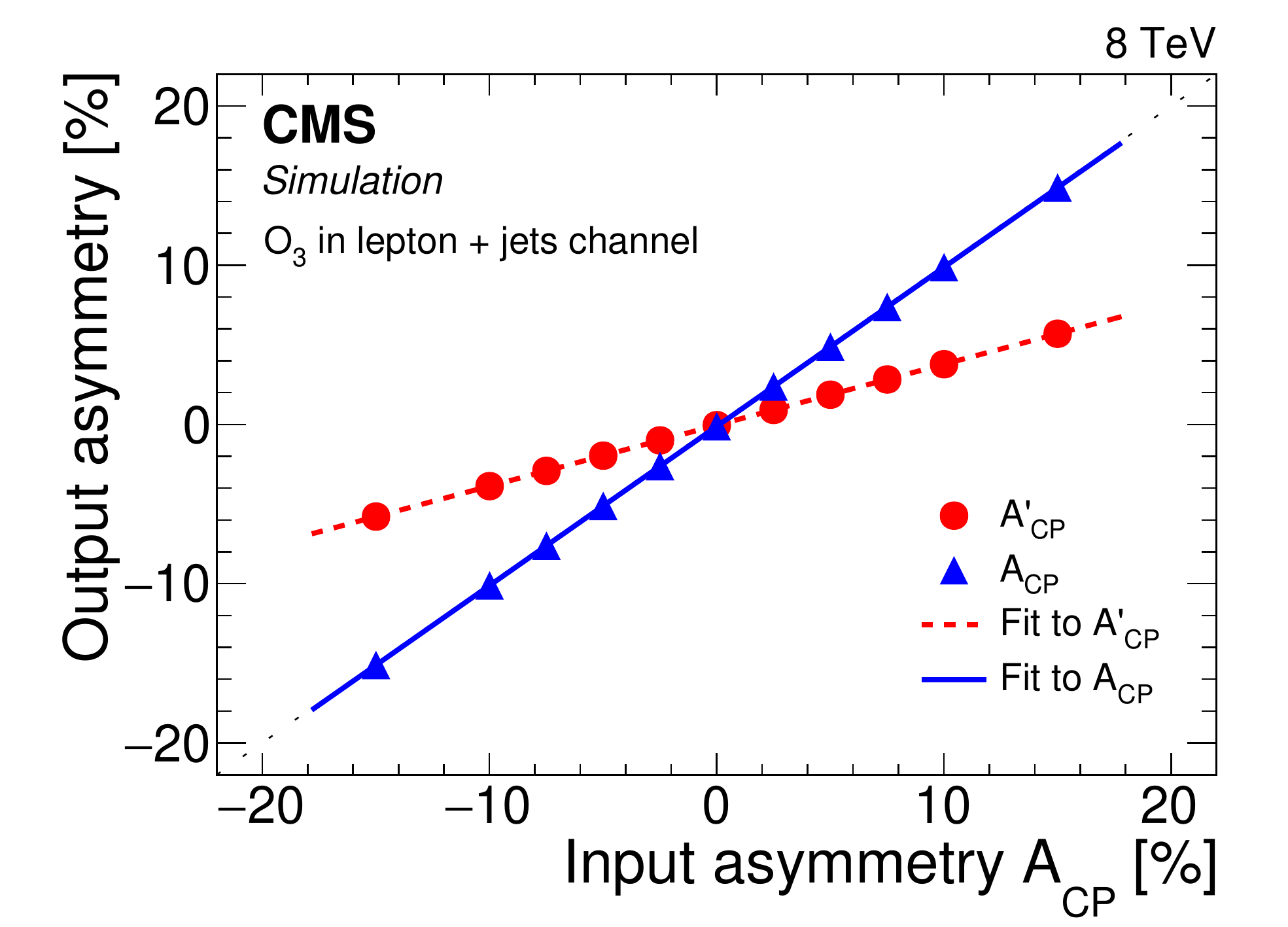}
    \includegraphics[width=0.42\textwidth]{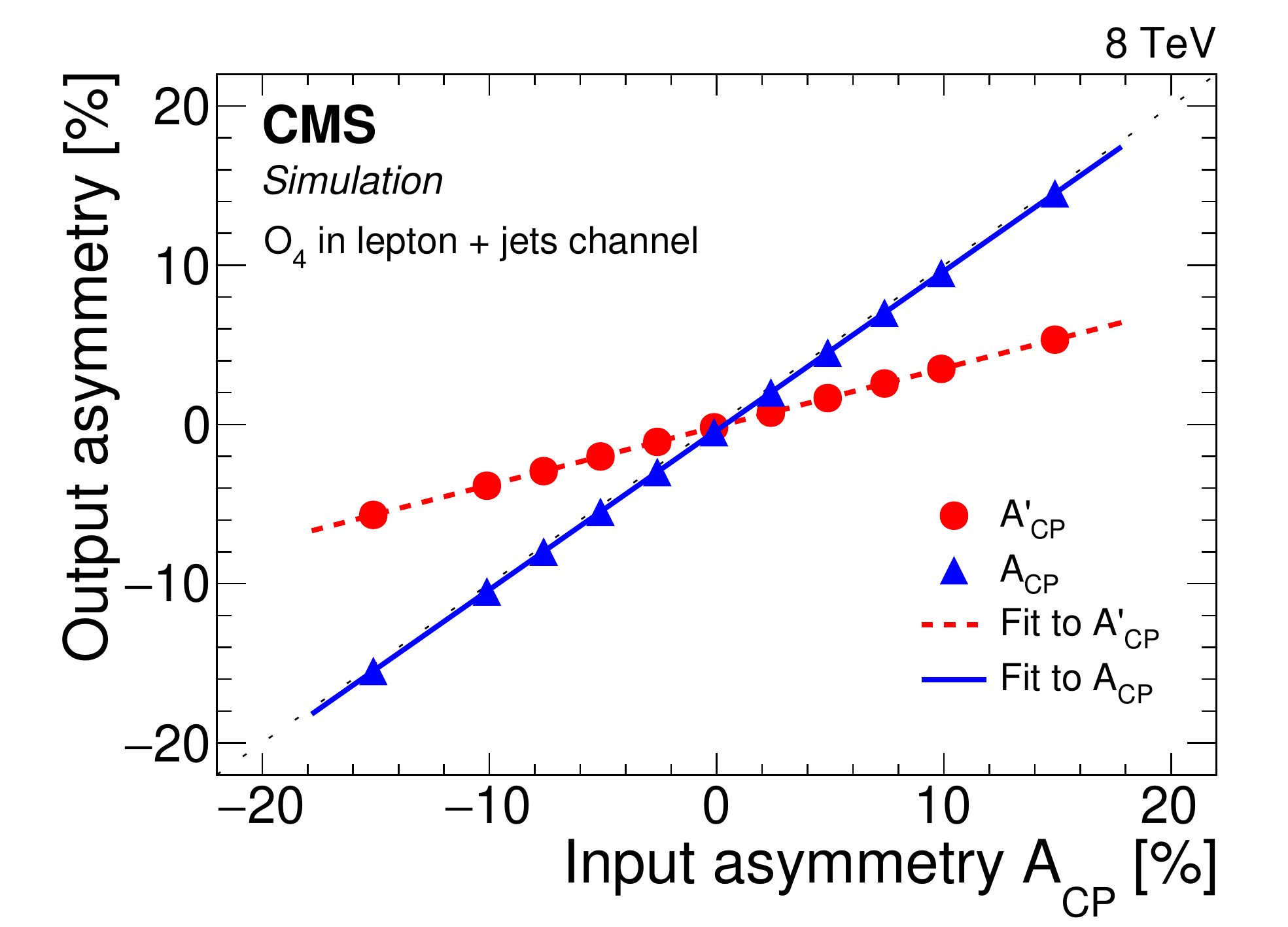}
    \includegraphics[width=0.42\textwidth]{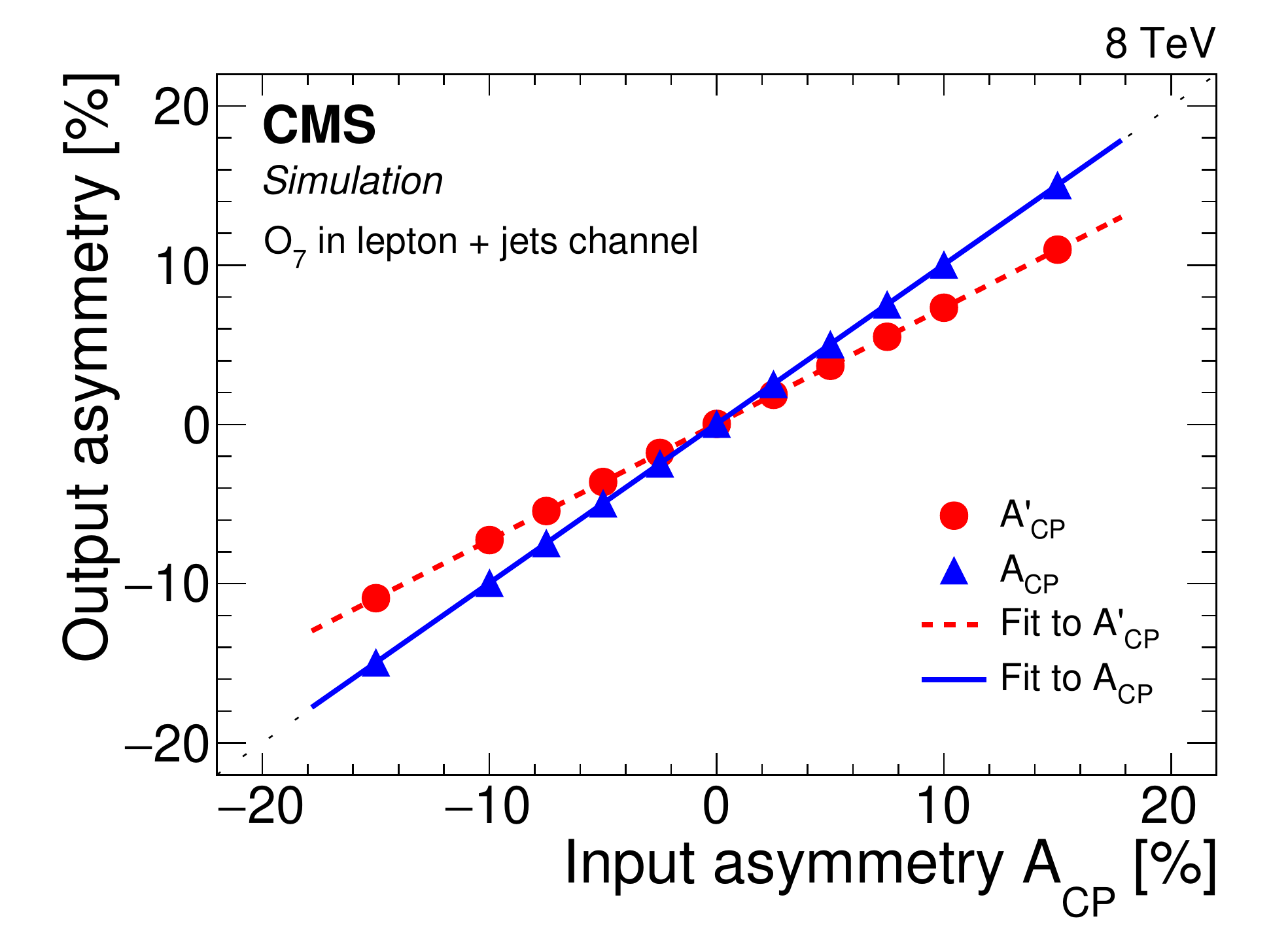}
    \caption{The results from simulation of the asymmetry correction procedure using a dilution factor for the four different CPV observables. The circular markers show the output \ACPp measurements for each generated \ACP value. The dashed lines are the result of linear fits to the \ACPp points. The triangular markers give the corrected \ACP values, obtained after applying the dilution factor. The solid lines are the result of linear fits to the corrected \ACP points. The statistical uncertainties in both sets of asymmetries are smaller than the markers.
    }
    \label{fig:ACP_dilute}
\end{figure}

To evaluate the sensitivity of the analysis to CPV, simulated events are reweighted at the generator level to produce hypothetical \ACP asymmetries in the observables. As shown in Fig.~\ref{fig:ACP_dilute}, the resulting values of \ACPp (circular points), extracted by treating the simulated events as data, exhibit a linear dependence on the generated \ACP values (dashed line). The \ACPp values are related to the generated \ACP values through dilution factors $\mathcal{D}$, applied as a multiplicative correction $\ACPp = \mathcal{D}\ACP$. The corrected asymmetries with the dilution factors applied (triangular points) are in agreement with the generator-level asymmetries, as shown by solid lines in Fig.~\ref{fig:ACP_dilute}, which are the results of fits to the corrected asymmetry values. The slopes of the lines are consistent with 1.0 and their $y$ intercepts consistent with 0.0.

As mentioned in the introduction, the dilution factors can be affected by new-physics processes. The reason for this is that the processes can alter the kinematic distributions of \ttbar events. For this reason, we consider the measurements of the uncorrected asymmetries \ACPp to be our primary result. It is nonetheless of interest to examine the overall size of the dilution factors, using the \ttbar simulation, for the cases when no new physics is present or when it is present to only a small degree.

The results for the dilution factors of the four observables are summarized in Table~\ref{tab:ACP_dilute}. If the sign of an observable at the generator level is the same as (different from) its sign in the reconstructed event, the event is classified as having the correct (wrong) sign. The fraction $k$ of wrong-sign events is related to the dilution factor that is defined by $\mathcal{D}=1-2k$.
 The value of $k$ for the four CPV observables are given in Table~\ref{tab:ACP_dilute}, along with their statistical and systematic uncertainties. The systematic uncertainties are estimated from the different conditions of correction and modeling, as mentioned in Section~\ref{sec:systematics}.
Studies reveal the following sources contribute to the value of $k$: misidentification of the b quark jet charge, 10.5\%; mistagging of b quark jets, 7.5\%; misassignment of the highest-\pt jet from W boson decay, 6.5\%; jet energy resolution, 2.5\% per jet. The effect of lepton charge misidentification is negligible.  The observables \Othree and \Ofour depend on similar quantities, including the charges of the b quark jets, and exhibit similar values of $k$. The observables \Otwo and \Oseven have lower $k$ values since fewer objects are used in their calculation.

\begin{table}[tb]
 \centering
  \topcaption{For each of the four CPV observables, the fraction $k$ of wrong-sign events and the associated dilution factor $\mathcal{D}$ computed from $k$, determined from simulated \ttbar events. The first uncertainty is statistical and the second is systematic.}
  \label{tab:ACP_dilute}
  \begin{tabular}{ccc}
   Observable &  Wrong-sign fraction $k$ (\%) & Dilution factor $\mathcal{D}$ \\
   \hline
   $\Otwo$   & $21.27 \pm 0.10 \pm 0.97$ & $0.575 \pm 0.002 \pm 0.019$ \\
   $\Othree$ & $30.86 \pm 0.10 \pm 0.90$ & $0.383 \pm 0.002 \pm 0.018$ \\
   $\Ofour$  & $31.65 \pm 0.10 \pm 0.95$ & $0.367 \pm 0.002 \pm 0.019$ \\
   $\Oseven$ & $13.52 \pm 0.11 \pm 0.50$ & $0.730 \pm 0.002 \pm 0.010$ \\
  \end{tabular}
\end{table}

\subsection{Asymmetry measurements}\label{ss:final_results}

\begin{figure}[tb]
  \centering
    \includegraphics[width=0.42\textwidth]{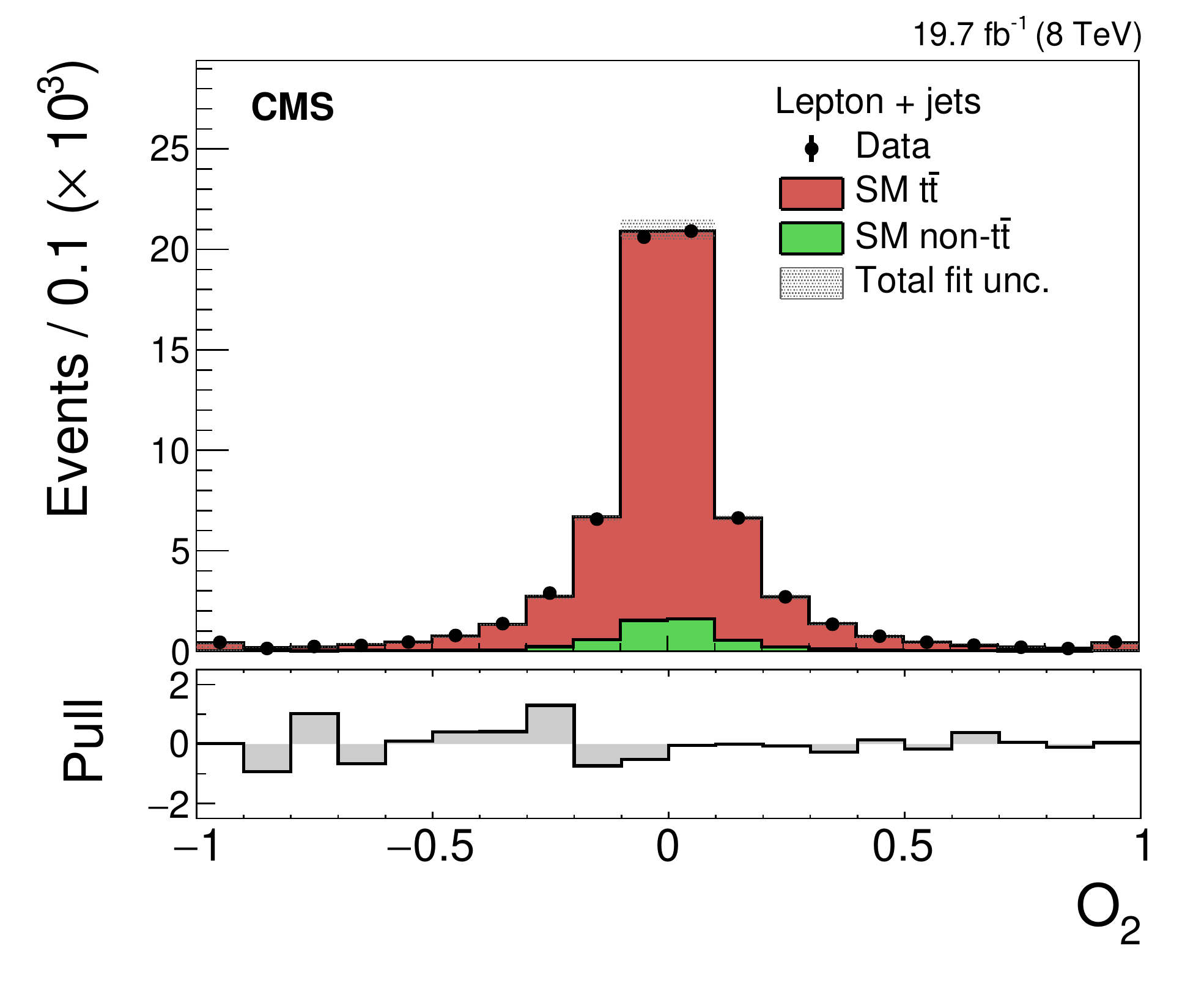}
    \includegraphics[width=0.42\textwidth]{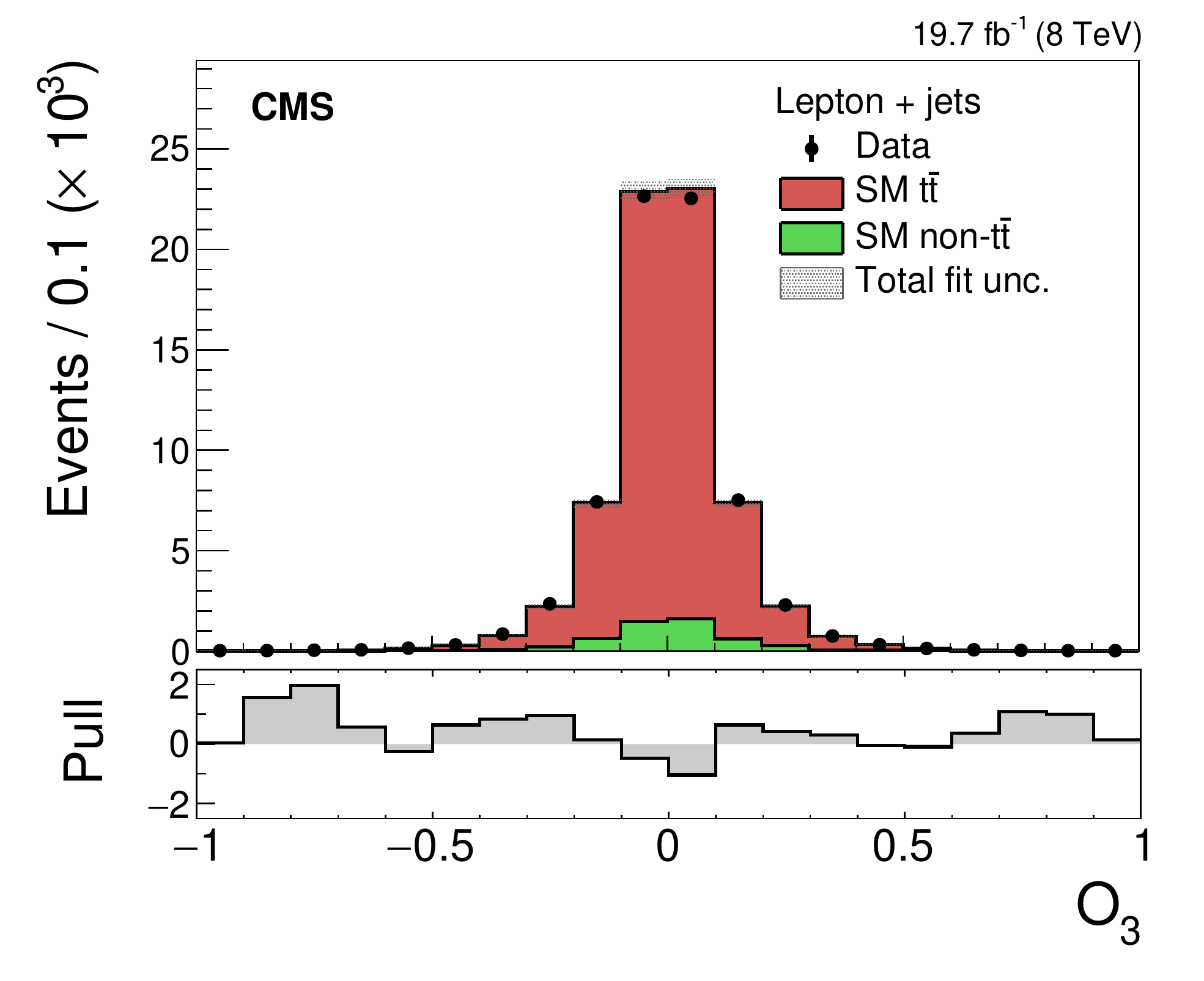}
    \includegraphics[width=0.42\textwidth]{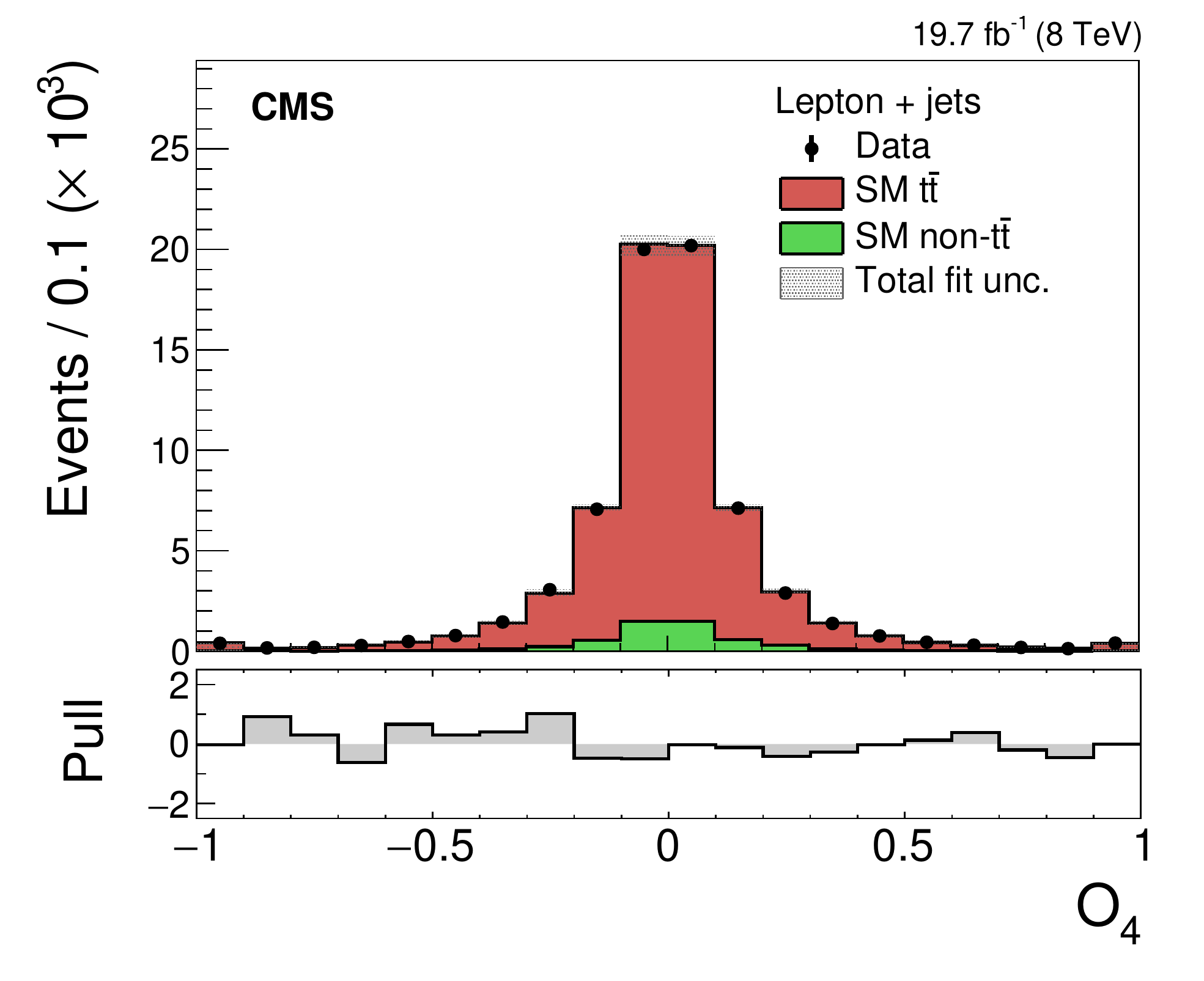}
    \includegraphics[width=0.42\textwidth]{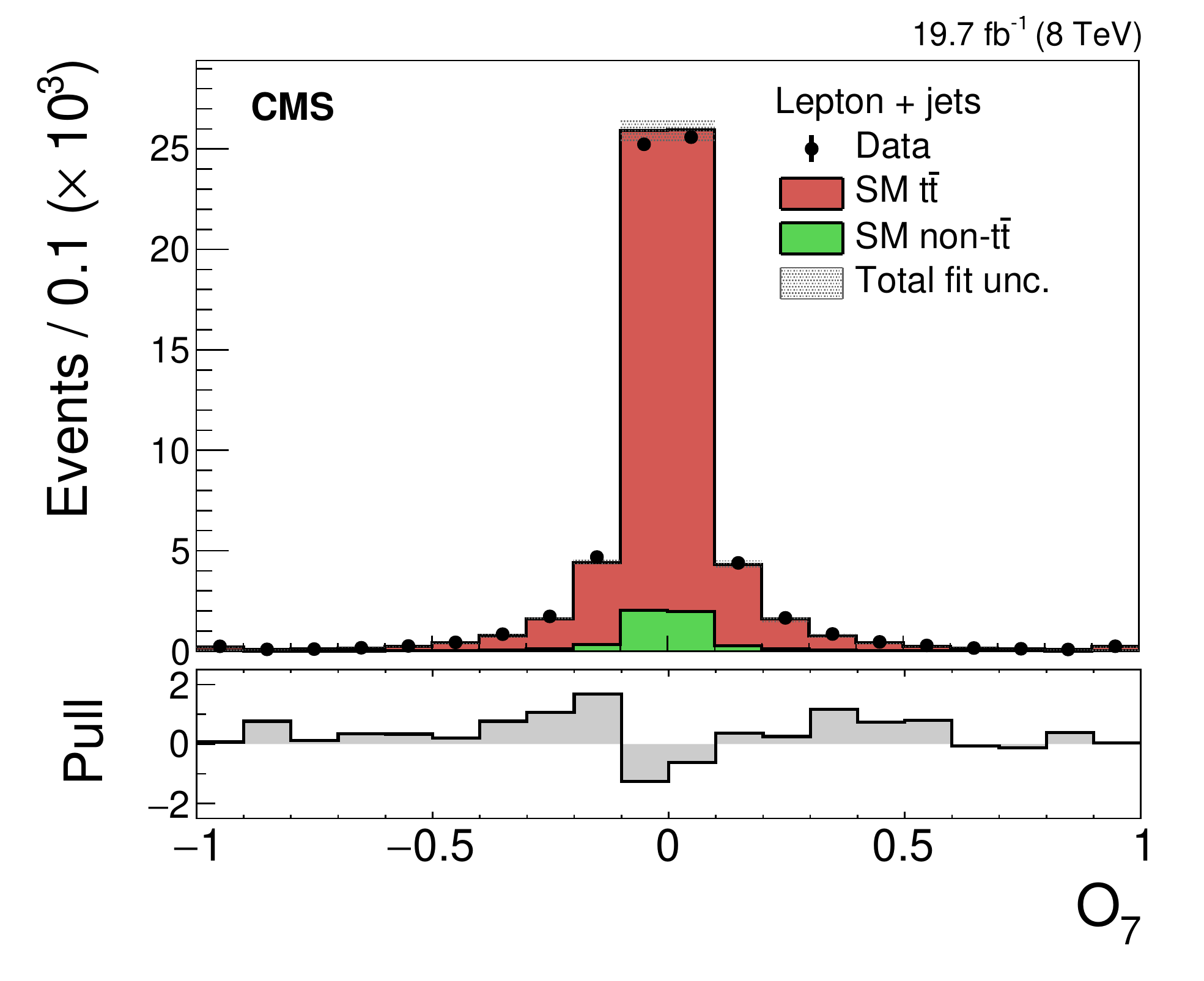}
  \topcaption{Distributions of the four CPV observables given in Eq.~(\ref{eq:defO}), determined from the combined electron and muon channels from data (points) and simulated signal and background (filled histograms). The simulated \ttbar and background samples are normalized to the fitted yields. The overflow events are collected in the first and last bins. Each observable is given in units of $m_{\PQt}^3$, where $m_{\PQt}=172.5\GeV$. The vertical bars represent the statistical uncertainties in the data. The hatched bands give the combined statistical and systematic uncertainties added in quadrature. The difference between the observed and expected events, divided by the total statistical and systematic uncertainty (pull), is shown for each bin in the lower panels.}
    \label{fig:O_unblind}
\end{figure}

\begin{table}[tb]
 \centering
  \topcaption{The uncorrected (corrected) CP asymmetry \ACPp (\ACP), measured in percent, for each of the four CPV observables. Results for \ACPp are given for the electron and muon channels separately and for their combination. For the \ACPp results, the first uncertainty is statistical and the second systematic. The \ACP values assume the dilution factors found from the SM simulation. The uncertainties in the \ACP results are the combined statistical and systematic terms added in quadrature.}
  \label{tab:ACP_unblind}
  \begin{tabular}{cccc|c}
   &  \multicolumn{3}{c|}{\ACPp (\%)} & \ACP (\%)\\
   &  e + jets & $\mu$ + jets & $\ell$ + jets & $\ell$ + jets\\
   \hline
   $\Otwo$    & $-0.19 \pm 0.61 \pm 0.59 $ & $+0.46 \pm 0.57 \pm 0.65$ & $+0.16 \pm 0.42 \pm 0.44$ & $+0.3 \pm 1.1 $\\
   $\Othree$  & $+0.02 \pm 0.61 \pm 0.59 $ & $-0.59 \pm 0.57 \pm 0.65$ & $-0.31 \pm 0.42 \pm 0.44$ & $-0.8 \pm 1.6 $\\
   $\Ofour$   & $-0.17 \pm 0.61 \pm 0.59 $ & $-0.10 \pm 0.57 \pm 0.65$ & $-0.13 \pm 0.42 \pm 0.44$ & $-0.4 \pm 1.7 $\\
   $\Oseven$  & $-0.38 \pm 0.61 \pm 0.59 $ & $+0.43 \pm 0.57 \pm 0.65$ & $+0.06 \pm 0.42 \pm 0.44$ & $+0.1 \pm 0.8 $\\
  \end{tabular}
\end{table}

\begin{figure}[htb]
  \centering
    \includegraphics[width=0.84\textwidth]{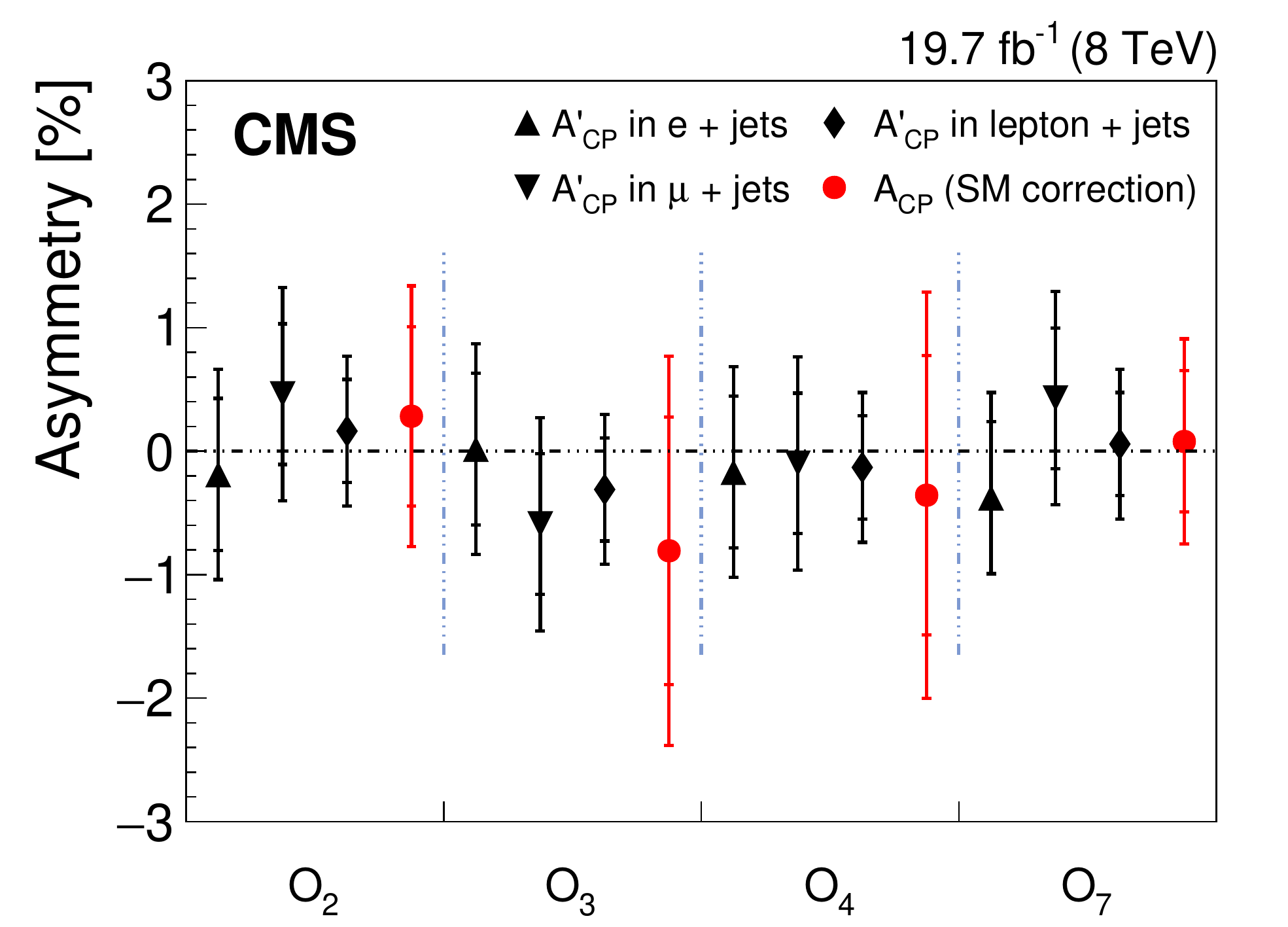}
    \caption{Summary of the uncorrected (corrected) CP asymmetries \ACPp (\ACP) for the observables defined in Eq.~(\ref{eq:defO}). The results for \ACPp are shown for the electron and muon channels separately and for their combination. The results for \ACP are shown for the combined electron and muon channels, using the dilution factors from SM simulation of \ttbar production. The inner bars represent the statistical uncertainties, and the outer bars the combined statistical and systematic uncertainties added in quadrature.
   }
    \label{fig:ACP_unblind}
\end{figure}

The measured distributions of the four observables are presented in Fig.~\ref{fig:O_unblind}. The corresponding \ACPp values, determined after subtraction of the background contributions, are shown in Table~\ref{tab:ACP_unblind} and displayed in Fig.~\ref{fig:ACP_unblind}. No significant nonzero asymmetry is observed in any of the separate electron or muon channels. Assuming that any new-physics process has at most a small effect on the \ttbar kinematic distributions, the dilution factors given in Table~\ref{tab:ACP_dilute} are applied to the combined sample to obtain the corrected \ACP values presented in Table~\ref{tab:ACP_unblind} and shown in Fig.~\ref{fig:ACP_unblind}.

\section{Summary}\label{sec:conclusion}

The first search for CP-violating effects in top quark-antiquark events has been presented. The search is performed in the electron + jets and muon + jets final states, with one top quark assumed to decay hadronically and the other semileptonically. The search is based on a sample of proton-proton collision data collected at $\sqrt{s}=8\TeV$ with the CMS detector in 2012, corresponding to an integrated luminosity of 19.7\fbinv.
 The CP-violating asymmetries are measured using four T-odd, triple-product observables, where T is the time-reversal operator.
 A data control sample is used to verify that no significant spurious CP asymmetry is introduced by background processes, and to model the shape of the background in the asymmetry observables.
 The normalization of the background contribution in the signal region is determined from a fit to the mass distribution $M_{\ell\PQb}$ associated with the semileptonically decaying top quarks. The background-subtracted distributions of the observables are used to compute the uncorrected asymmetries. The corrected asymmetries are obtained by using a multiplicative dilution factor derived from simulation. Both the uncorrected and corrected asymmetries are consistent with zero, in agreement with the expectation from the standard model.

\begin{acknowledgments}
We congratulate our colleagues in the CERN accelerator departments for the excellent performance of the LHC and thank the technical and administrative staffs at CERN and at other CMS institutes for their contributions to the success of the CMS effort. In addition, we gratefully acknowledge the computing centers and personnel of the Worldwide LHC Computing Grid for delivering so effectively the computing infrastructure essential to our analyses. Finally, we acknowledge the enduring support for the construction and operation of the LHC and the CMS detector provided by the following funding agencies: BMWFW and FWF (Austria); FNRS and FWO (Belgium); CNPq, CAPES, FAPERJ, and FAPESP (Brazil); MES (Bulgaria); CERN; CAS, MoST, and NSFC (China); COLCIENCIAS (Colombia); MSES and CSF (Croatia); RPF (Cyprus); SENESCYT (Ecuador); MoER, ERC IUT, and ERDF (Estonia); Academy of Finland, MEC, and HIP (Finland); CEA and CNRS/IN2P3 (France); BMBF, DFG, and HGF (Germany); GSRT (Greece); OTKA and NIH (Hungary); DAE and DST (India); IPM (Iran); SFI (Ireland); INFN (Italy); MSIP and NRF (Republic of Korea); LAS (Lithuania); MOE and UM (Malaysia); BUAP, CINVESTAV, CONACYT, LNS, SEP, and UASLP-FAI (Mexico); MBIE (New Zealand); PAEC (Pakistan); MSHE and NSC (Poland); FCT (Portugal); JINR (Dubna); MON, RosAtom, RAS, and RFBR (Russia); MESTD (Serbia); SEIDI and CPAN (Spain); Swiss Funding Agencies (Switzerland); MST (Taipei); ThEPCenter, IPST, STAR, and NSTDA (Thailand); TUBITAK and TAEK (Turkey); NASU and SFFR (Ukraine); STFC (United Kingdom); DOE and NSF (USA).

\hyphenation{Rachada-pisek} Individuals have received support from the Marie-Curie program and the European Research Council and EPLANET (European Union); the Leventis Foundation; the A. P. Sloan Foundation; the Alexander von Humboldt Foundation; the Belgian Federal Science Policy Office; the Fonds pour la Formation \`a la Recherche dans l'Industrie et dans l'Agriculture (FRIA-Belgium); the Agentschap voor Innovatie door Wetenschap en Technologie (IWT-Belgium); the Ministry of Education, Youth and Sports (MEYS) of the Czech Republic; the Council of Science and Industrial Research, India; the HOMING PLUS program of the Foundation for Polish Science, cofinanced from European Union, Regional Development Fund, the Mobility Plus program of the Ministry of Science and Higher Education, the National Science Center (Poland), contracts Harmonia 2014/14/M/ST2/00428, Opus 2014/13/B/ST2/02543, 2014/15/B/ST2/03998, and 2015/19/B/ST2/02861, Sonata-bis 2012/07/E/ST2/01406; the Thalis and Aristeia programs cofinanced by EU-ESF and the Greek NSRF; the National Priorities Research Program by Qatar National Research Fund; the Programa Clar\'in-COFUND del Principado de Asturias; the Rachadapisek Sompot Fund for Postdoctoral Fellowship, Chulalongkorn University and the Chulalongkorn Academic into Its 2nd Century Project Advancement Project (Thailand); and the Welch Foundation, contract C-1845.
\end{acknowledgments}

\bibliography{auto_generated}

\cleardoublepage \appendix\section{The CMS Collaboration \label{app:collab}}\begin{sloppypar}\hyphenpenalty=5000\widowpenalty=500\clubpenalty=5000\input{TOP-16-001-authorlist.tex}\end{sloppypar}
\end{document}

%% file: TOP-16-001-authorlist.tex
\textbf{Yerevan Physics Institute,  Yerevan,  Armenia}\\*[0pt]
V.~Khachatryan, A.M.~Sirunyan, A.~Tumasyan
\vskip\cmsinstskip
\textbf{Institut f\"{u}r Hochenergiephysik,  Wien,  Austria}\\*[0pt]
W.~Adam, E.~Asilar, T.~Bergauer, J.~Brandstetter, E.~Brondolin, M.~Dragicevic, J.~Er\"{o}, M.~Flechl, M.~Friedl, R.~Fr\"{u}hwirth\cmsAuthorMark{1}, V.M.~Ghete, C.~Hartl, N.~H\"{o}rmann, J.~Hrubec, M.~Jeitler\cmsAuthorMark{1}, A.~K\"{o}nig, I.~Kr\"{a}tschmer, D.~Liko, T.~Matsushita, I.~Mikulec, D.~Rabady, N.~Rad, B.~Rahbaran, H.~Rohringer, J.~Schieck\cmsAuthorMark{1}, J.~Strauss, W.~Waltenberger, C.-E.~Wulz\cmsAuthorMark{1}
% ...
\vskip\cmsinstskip
% ...
1:~~Also at Vienna University of Technology, Vienna, Austria\\
% ...
